\documentclass[nofootinbib,twocolumn,aps,prd,amsmath,superscriptaddress]{revtex4-1}
\usepackage{setspace}
\usepackage{fancyhdr}
\usepackage{graphicx}
\usepackage[ansinew]{inputenc}
\usepackage{amssymb}
\usepackage{amsmath}
\usepackage{float}
\usepackage{subfigure}
\usepackage{hyperref}

\newcommand{\etal}{\textit{et al.}}

\renewcommand{\epsilon}{\varepsilon}

\newcommand{\mnuc}{m_n}

\newcommand{\yq}{Y_q}

\newcommand{\nb}{n_b}

\newcommand{\nii}{n_i}
\newcommand{\s}{s/\nb}
\newcommand{\szero}{s_0/\nb}

\newcommand{\mub}{\mu_b}

\newcommand{\mul}{\mu_\ell}
\newcommand{\mui}{\mu_i}

\newcommand{\munu}{\mu_\nu}
\newcommand{\mgrav}{M_G}
\newcommand{\mbary}{M_B}

\newcommand{\kb}{k_B}
\newcommand{\kbperbaryon}{\kb/\mathrm{baryon}}
\newcommand{\msol}{\mathrm{M}_\odot}

\newcommand{\gpercmcub}{\mathrm{g/cm}^3}

\newcommand{\mev}{\mathrm{MeV}}
\newcommand{\mevperfmcub}{\mathrm{MeV/fm}^3}

\newcommand{\catitoiv}{\mathrm{I} - \mathrm{IV}}
\newcommand{\catdc}{\mathrm{D} \rightarrow \mathrm{C}}
\newcommand{\catad}{\mathrm{A} \rightarrow \mathrm{D}}
\newcommand{\catadc}{\mathrm{A} \rightarrow \mathrm{D} \rightarrow \mathrm{C}}
\newcommand{\catac}{\mathrm{A} \rightarrow \mathrm{C}}

\newcommand{\epshm}{\epsilon_{\mathrm{HM}}}
\newcommand{\epstrans}{\epsilon_{\mathrm{trans}}}
\newcommand{\ptrans}{p_{\mathrm{trans}}}
\newcommand{\mutrans}{\mu_{\mathrm{trans}}}
\newcommand{\ttrans}{T_{\mathrm{trans}}}
\newcommand{\ntranshm}{n_{\mathrm{trans}}^{\mathrm{HM}}}
\newcommand{\ntransqm}{n_{\mathrm{trans}}^{\mathrm{QM}}}

\newcommand{\cqm}{c_{\mathrm{QM}}}
\newcommand{\epscrit}{\epsilon_{\mathrm{crit}}}

\newcommand{\nzero}{n_0}

\begin{document}
	\title{Twins in the Shadow: Phase Transitions in Proto-Neutron Stars}
    \author{Jannik Milthaler}
	\affiliation{Hamburger Sternwarte, University of Hamburg, Gojenbergsweg 112, 21029 Hamburg, Germany}
    \author{Jan-Erik Christian}
	\email{jan-erik.christian@uni-hamburg.de}
	\affiliation{Hamburger Sternwarte, University of Hamburg, Gojenbergsweg 112, 21029 Hamburg, Germany}
	\author{Pia Jakobus}
	\affiliation{Hamburger Sternwarte, University of Hamburg, Gojenbergsweg 112, 21029 Hamburg, Germany}
	\author{Stephan Rosswog}
	\email{stephan.rosswog@uni-hamburg.de}
	\affiliation{Hamburger Sternwarte, University of Hamburg, Gojenbergsweg 112, 21029 Hamburg, Germany}
	\affiliation{The Oskar Klein Centre, Department of Astronomy, AlbaNova, Stockholm University, SE-106 91 Stockholm, Sweden}
    
	\date{\today}
	\begin{abstract}
    We study the impact of a first-order phase transition from hadronic to deconfined quark matter on the evolution of proto-neutron stars. To maximize the applicability of our results, we construct an equation of state that is highly parameterizable, where we use a relativistic mean field approach to model hadronic matter and a constant speed of sound approach to model quark matter. We approximate the cooling behavior of proto-neutron stars as a sequence of decreasing entropy at constant baryonic mass.
    Some parameterizations lead to so-called "twin-stars", where two neutron stars have the same (gravitational) mass but different radii. Detecting them is regarded as a smoking gun signal for phase transition. We find, however, that proto-neutron star evolution suggests that gravitational twin stars are only possible if the hybrid star does not have, in addition, a baryonic twin. This rules out most twin star pairs, especially if the phase transition occurs at large densities. In a mass-radius diagram the unpopulated hybrid branch appears to be ``in the shadow'' of the hadronic branch.
	\end{abstract}
	\maketitle

	\section{Introduction}\label{incanus}
   Proto-neutron stars (PNSs) are the immediate remnants formed during the core collapse of massive stars. During the collapse of the stellar core the density can surpass multiple times nuclear saturation density $\rho_0\simeq2.7\times10^{14}\,\mathrm{g\,cm^{-3}}$, which leads to a stiffening of the equation of state (EoS) due to repulsive short-range nuclear interactions and turns the infall into an explosion. This ``bounce'' of the core generates a shock wave that initially propagates outward but loses large amounts of its kinetic energy through photo-dissociation of heavy nuclei and neutrino losses in the outer core. Several hundred milliseconds after core bounce, the shock wave may be revived through neutrino heating in the region behind the stalled shock, a process known as the delayed neutrino heating mechanism, which is the standard paradigm for core-collapse supernova explosions~\cite{Colgate:1966,Bethe:1985}, see \cite{Janka:2006fh,Foglizzo:2015,Burrows:2021,Muller:2020ard} for reviews on the topic. During the early stage of the explosion, the PNS is extremely hot ($T \gtrsim 10$ MeV), lepton-rich, and rapidly evolving. Over timescales of tens of seconds, it cools and deleptonizes primarily through intense neutrino emission, eventually settling into a cold, stable neutron star~\cite{Glendenning:1994za,Keil:1995hw,Prakash:1997,Pons:1999,Prakash:2001,Dexheimer:2008ax,Roberts:2012,Camelio:2017,Pascal:2022}.
   \\\\
    Due to their high central densities, neutron stars are ideal investigatory objects to probe the EoS under extreme conditions. One interesting prospect is the onset of a phase transition from hadronic to quark matter somewhere within the density range of a neutron star core. Such a configuration is called a hybrid star and it features a quark matter core together with a hadronic mantle \cite{Ivanenko:1965dg,Itoh:1970uw,Alford:2004pf,Coelho:2010fv,Chen:2011my,Masuda:2012kf,Yasutake:2014oxa,Zacchi:2015oma,Xie:2020rwg}. Hybrid stars cannot easily be distinguished from purely hadronic neutron stars, but, under some circumstances, a phase transition could change the mass-radius relation drastically enough so that two stars with the same mass but different radii are produced. Such stars are referred to as twin stars, and they are a nearly unique indicator for the presence of a phase transition \cite{Kampfer:1981yr,Glendenning:1998ag,Schertler:2000xq,SchaffnerBielich:2002ki,Zdunik:2012dj,Blaschke:2015uva,Zacchi:2016tjw,Christian:2018jyd,Blaschke:2019tbh,Tan:2020ics}.
    \\\\
    In this work, we approximate the cooling phase of PNSs by tracking the baryonic masses of individual stars through phases of decreasing, uniform entropy, and we examine the influence of a phase transition on the PNS evolution. We use the constant speed of sound (CSS) parametrization~\cite{Zdunik:2012dj,Alford:2013aca,Alford:2015dpa,Alford:2017qgh,Christian:2017jni,Rau:2022ofy,Christian:2023hez,Li:2024lmd} to describe quark matter. While this approach does not contain any substantial microphysical information, it allows us to make generalized statements about the occurrence of phase transitions without a strong model dependence. In fact, even though we refer to the CSS parameterized phase as quark matter, other exotic types of matter preceded by a sufficiently strong phase transition are equally applicable. 
    \\\\
   In section~\ref{sec:framework}, we present our framework to approximate the PNS evolution. This is followed by a short analysis of purely hadronic EoSs within this framework in section~\ref{sec:results_without_qm}, allowing us to compare our results with a phase transition in section~\ref{sec:results_with_qm}. In section~\ref{sec:css_categorization} we discuss the constraints that our approach imposes on the parameter space of hybrid stars. Finally we summarize and discuss our results in section~\ref{conclusion}.
    
    \section{Theoretical Framework}
        \label{sec:framework}
    \subsection{Description of Matter}
     \label{sec:eos}
        \subsubsection{Hadronic Matter}
    To explore the effect of a phase transition on the evolution of PNSs, we investigate a plethora of hadronic EoSs in section \ref{sec:results_without_qm} before we include quark matter. The EoSs we employ are SRO(SLy4)~\cite{website:SLy4,Schneider:2017,Lattimer:1991nc,Chabanat:1998}, SRO(APR)~\cite{website:APR,Akmal:1998cf,Schneider:2017,Schneider:2019}, HS(DD2)~\cite{website:DD2,Hempel:2010,Typel:2009sy,Moeller:1997}, HS(IUF)~\cite{website:IUF,Hempel:2010,Fattoyev:2010,Roca-Maza:2008}, and SFH(SFHo)~\cite{website:SFHo,Hempel:2010,Moeller:1997,Steiner:2013}, which we obtained from the CompOSE website~\cite{website:compose,Typel:2015,Oertel:2016bki,Typel:2022}.
    In addition, we use multiple parameter sets of the Frankfurt-Barcelona (FB) EoS, which is a finite temperature extension \cite{Christian:2023byx} of the relativistic mean field model used by Hornick \etal~\cite{Hornick:2018kfi}. In this model the symmetry energy $J$, slope parameter $L$ and effective nucleon mass $m^*$ are varied within the limits set by chiral effective field theory predictions and nuclear constraints. Here we set $J=32\,\mev$ and $L=60\,\mev$, which maximizes the range of possible $m^*$. This approach allows us to connect in a consistent way the effective mass to the EoS's stiffness, where smaller effective masses correspond to stiffer EoSs~\cite{Boguta:1981px}.
    We use the values $m^*/m_n= 0.65$, $0.70$, and $0.75$, where $m_n$ refers to the nucleon mass. A value of $m^*/m_n = 0.65$ leads to the stiffest FB EoS that is still compatible with GW170817~\cite{Hornick:2018kfi}, while $m^{*}/m_n = 0.75$ yields the softest FB EoS that can still reach $2\,M_\odot$. We refer to these EoSs as FB65, FB70, and FB75. In addition, we include one case where we change the slope parameter to $L=40\,\mev$, which we denote as FB70L40.\\
    In section \ref{sec:pns_evolution_with_qm} we focus on the impact of phase transitions, we therefore limit ourselves to only one EoS. We chose the FB65 because it is the stiffest EoS in the FB family that is compatible with astrophysical results, provided there is no phase transition.\\
    The FB EoS only describes bulk matter, therefore, we attach a low-density region described by the HS EoS~\cite{Hempel:2010}.
    In some cases our tables are insufficient in the small entropy and density regime. In such cases we switch to the zero temperature BPS EoS~\cite{Baym:1971pw,Negele:1971vb}.
    
        \subsubsection{Phase Transition and Quark Matter}
    \label{sec:css_parametrization}
    For the quark matter description, we use a constant speed of sound (CSS) parametrization in combination with a Maxwell construction \cite{Zdunik:2012dj,Alford:2013aca}, which lets us express the energy density $\epsilon$ as a function of pressure $p$:
    \begin{equation}\label{Maxwell}
    	\epsilon(p) =
    	\begin{cases}
    		\epshm(p) & \ \ p < \ptrans \\
    		\epstrans + \Delta\epsilon + \cqm^{-2} (p - \ptrans) & \ \ p > \ptrans.
    	\end{cases}
    \end{equation}
    This approach has three free parameters: the transition pressure, $\ptrans$, the jump in energy density at the point of transition, $\Delta\epsilon$, and the (constant) speed of sound in the quark phase, $\cqm$. Like in previous studies, we fix $\cqm=1$ in natural units, see  e.g.~\cite{Christian:2023hez}. The main advantage of the CSS approach lies in its generalizability~\cite{Alford:2013aca}, but it does not provide microphysical information on the quark phase. In the following, the baryon number density, $\nb$, is required to investigate PNS behavior. We approximate $\nb$ in the quark phase following Drischler \etal~\cite{Drischler:2020fvz} as
    \begin{equation}
    	\nb = \ntransqm \left(\frac{\epsilon + p}{\epstrans + \Delta\epsilon + \ptrans}\right)^{1/(1+\cqm^2)}\,{.}
    \end{equation}
    Using \eqref{Maxwell}, we can solve this for $p(\nb)$ in the quark phase:
    \begin{equation}
    	\begin{split}
    		p(\nb) = \left(\frac{\nb}{\ntransqm}\right)^{1 + \cqm^2} \, \frac{\epstrans + \Delta\epsilon + \ptrans}{1 + \cqm^{-2}} \\
    		+ \frac{\cqm^{-2}\,\ptrans - \epstrans - \Delta\epsilon}{1 + \cqm^{-2}}\,{.}
    	\end{split}
    \end{equation}
    In a Maxwell construction, $p$, $\mub$ and $T$ are constant, while $\nb$ and $\epsilon$ increase from $\ntranshm$ and $\epstrans$ in the hadronic phase to $\ntransqm$ and $\epstrans + \Delta\epsilon$ respectively. The value of $\ntransqm$ can be calculated from the Euler equation~\cite{1959flme.book.....L}:
    \begin{equation}
    	\epsilon = -p + T s + \sum_i \mui \nii\,{,}
    \end{equation}
    which yields
    \begin{equation}
    	\ntransqm = \frac{\epstrans + \Delta\epsilon + \ptrans}{\ttrans\,\s + \mutrans}\,{.}
    \end{equation}

    \subsection{Proto-Neutron Star Evolution}\label{subsec:PNS_Evolution}
    To approximate the evolution of a PNS, we use a quasi-static approach, in which we take multiple snapshots at different cooling stages. This relies on the assumption that the timescale to establish hydrostatic equilibrium is much shorter than the neutrino cooling timescale~\cite{Pons:1999,Pascal:2022}.
    For each snapshot, we calculate the mass-radius (MR) relation using the Tolman-Oppenheimer-Volkoff (TOV) equations~\cite{Tolman39,OV39}. Since the used EoS tables end at finite temperatures, we stop the integration at a mass density of $\rho = 10^{11}\,\gpercmcub$, which is usually used to define the PNS boundary in core-collapse supernovae (CCSNe) simulations~\cite{Burrows:2019,Nagakura:2020,Jakobus:2022}.
    We track the PNS evolution as a sequence of decreasing, uniform entropy at fixed baryonic mass.
    Compared to the gravitational mass
    \begin{equation}
        \mgrav = \int_0^R 4 \pi r^2 \,\epsilon(r) \,dr\ ,
    \end{equation}
    the baryonic mass
    \begin{equation}
        \mbary = \int_0^R \frac{4 \pi r^2 \,\mnuc \nb(r)}{\left[1 - 2 \,G \,m(r) / r\right]^{1/2}} \,dr\,{,}
    \end{equation}
    is calculated by integrating the baryon number density $\nb$ multiplied by the nucleon mass $\mnuc$ excluding the gravitational binding energy~\cite{Glendenning:2000,Bombaci:1996}. In essence, this is equivalent to summing up the masses of the total number of baryons.
    Therefore, a constant baryonic mass implies no accretion onto the PNS. This is a reasonable assumption, because mass accretion ceases quickly after core bounce on the order of a second~\cite{Pons:1999,Burrows:2019,Nagakura:2020}.
    \\\\
    Because convection leads to nearly flat profiles of entropy $\s$ and charge fraction $\yq$ for large parts of the PNS~\cite{Roberts:2012,Nagakura:2020,Pascal:2022,Jakobus:2022,Jakobus:2025}, we employ for simplicity completely flat profiles of entropy and charge fraction for the whole PNS. 
    We investigate a range of entropies from ${\s=3.0\,\kbperbaryon}$ down to ${\s=0.1\,\kbperbaryon}$ in steps of 0.1 to mimic the PNS cooling behavior.
    For the entropy profile this is the usual approach, however, for the charge fraction numerous approximations are in use, like the assumption of beta equilibrium or a flat profile of the lepton fraction~\cite{Prakash:1997,Pons:1999,Dexheimer:2008ax,Camelio:2017,Hempel:2017,Mariani:2017}. We assume a constant charge fraction for the entire PNS evolution, because this allows us to take a large number of snapshots without the need to construct an entropy dependence for $\yq$, which would require further assumptions. This approximation has some drawbacks, which are outlined in section \ref{sec:results_without_qm}.
    \\\\
    Initially, a PNS contains a significant amount of trapped neutrinos, which slowly diffuse out of the star as it cools down until it finally becomes transparent~\cite{Keil:1995hw,Pons:1999,Prakash:1997,Prakash:2001,Camelio:2017}. We approximate the contributions from neutrinos with a simple approach: During most of the PNS evolution, neutrinos are in thermal and chemical equilibrium  with matter and follow a Fermi-Dirac distribution~\cite{Prakash:2001}. Therefore, we treat them as an ideal ultra-relativistic Fermi gas~\cite{Hempel:2009vp,Greiner:1995book} and calculate their energy density, pressure, and entropy density as \cite{Greiner:1995book}
    \begin{equation}
        \epsilon_\nu = \frac{(\kb T)^4}{(\hbar c)^3} \left[ \frac{7\pi^2}{120} + \frac{1}{4}\left(\frac{\mul}{\kb T}\right)^2 + \frac{1}{8\pi^2}\left(\frac{\mul}{\kb T}\right)^4 \right]\,{,}
    \end{equation}
    \begin{equation}
        p_\nu = \frac{1}{3}\epsilon_\nu\,{,}
    \end{equation}
    and
    \begin{equation}
        \frac{S_\nu}{V} = \kb\left(\frac{\kb T}{\hbar c}\right)^3 \left[ \frac{7\pi^2}{90} + \frac{1}{6}\left(\frac{\mul}{\kb T}\right)^2 \right]\,{.}
    \end{equation}
    This allows us to approximate the neutrino contribution solely using the temperature $T$ and neutrino chemical potential $\munu$, which corresponds to the lepton chemical potential $\mul$ provided by the EoS tables, where $\mul$ is the sum of the chemical potentials of the included lepton-species.
    \\\\
    The final approximation concerns the entropy dependence of the phase transition. At low densities, the hadron-quark phase boundary is located at a constant energy density~\cite{Bazavov:2019}. We therefore fix the energy density at the point of transition:
    \begin{equation}
    	\epstrans(s)=\epstrans(\ptrans(s=0))=\epstrans,
    \end{equation}
    which allows us to approximate $\ptrans(s)$ by choosing a value that reproduces $\epstrans$. 
    
    \subsection{Hybrid and Twin Stars}
    \label{sec:hybrids_and_twins}
    \begin{figure}
    	\centering
    	\includegraphics[width=0.45\textwidth,keepaspectratio]{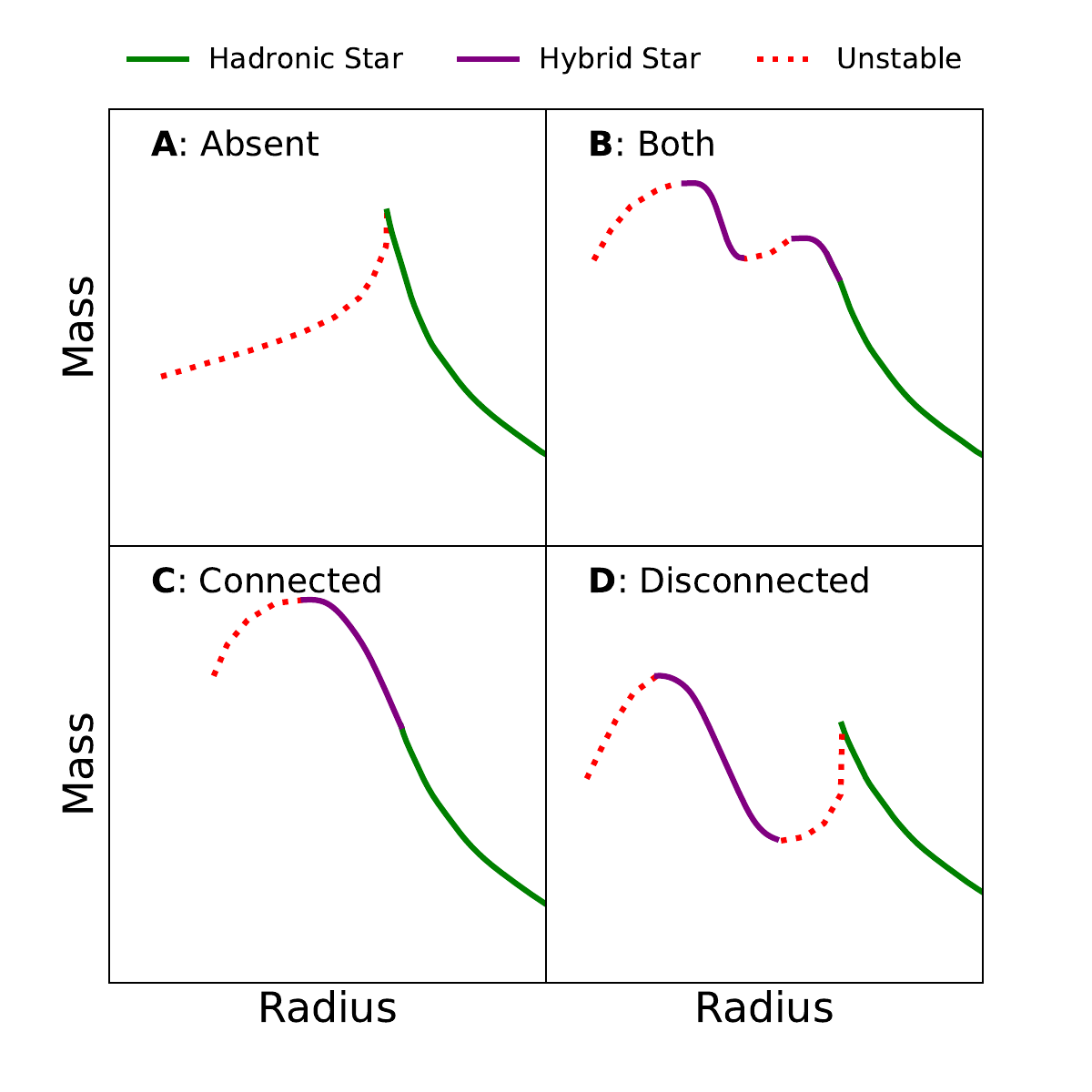}
    	\caption{Classification scheme of hybrid star EoSs adapted from Alford \etal~\cite{Alford:2013aca}. Purely hadronic stars are shown as green, hybrid stars are shown as purple lines and the unstable regions are shown as red dotted lines.
       }
    	\label{fig:Alford_Cats}
    \end{figure}
        \begin{figure}
    	\centering
    	\includegraphics[width=0.40\textwidth,keepaspectratio]{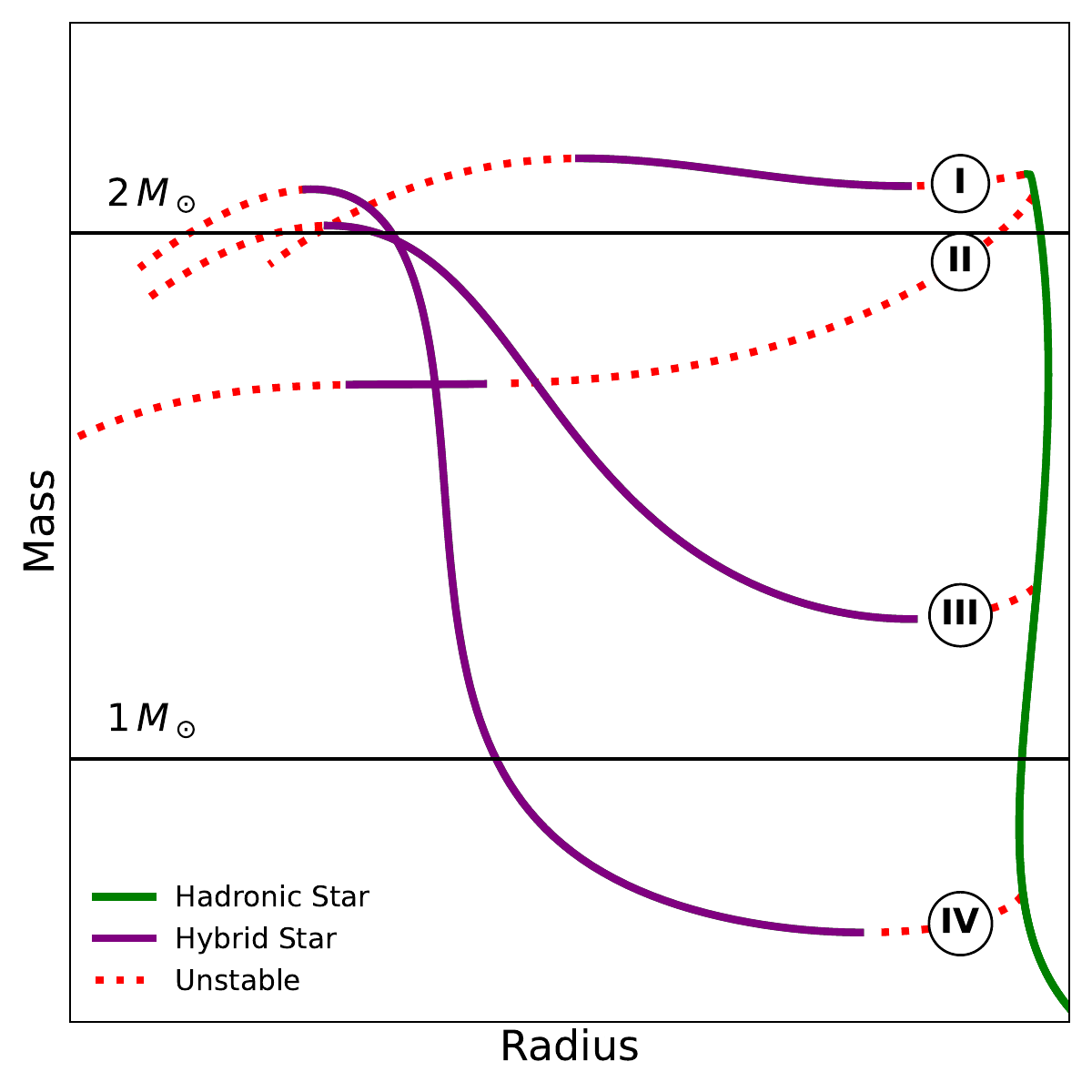}
    	\caption{Classification scheme for twin stars adapted from Christian \etal~\cite{Christian:2017jni}. 
        }
    	\label{fig:Christian_Cats}
    \end{figure} 
    Introducing a phase transition to the EoS changes the mass-radius relation once the central density increases beyond the transition point. Stars with higher central densities than the transition density are referred to as hybrid stars, which feature a quark matter core and a hadronic outer part. We chose the generic approach outlined above in order to make generalized statements about the impact of such configurations. This strategy has been employed, perhaps most prominently, by Alford \etal~\cite{Alford:2013aca} who used it to classify hybrid EoSs into four separate cases. 
    The quantity on which their classification scheme relies is the so-called Seidov limit~\cite{Seidov:1971}, which provides a critical value of $\Delta\epsilon$:
       \begin{equation}
    	\Delta\epscrit = \epstrans \left(\frac{1}{2} + \frac{3}{2}\frac{\ptrans}{\epstrans}\right)\,{.}
    \end{equation}
    
    Exceeding this limit corresponds to a destabilization of the mass-radius relation at the transition point. However, even if $\Delta\epsilon<\Delta\epscrit$ the sequence might become unstable shortly after the transition density is reached. In either case, the mass-radius relation can regain stability at higher central density, forming a second branch entirely composed of hybrid stars. 
    Based on this, Alford \etal~\cite{Alford:2013aca} find that hybrid stars are either absent (A) from the mass-radius relation, connected (C) to the first branch, disconnected (D) from the first branch, or present in both (B) branches. In the latter case, the hybrid part of the first branch is exceedingly small~\cite{Alford:2013aca,Christian:2017jni}. For this reason, we will not distinguish between cases B and D later in this work. We provide a schematic depiction of this ABCD-categorization in Fig. \ref{fig:Alford_Cats}.\\
    Furthermore, discretization errors (step size in TOV solver) can lead to hybrid stars that appear to be stable in the first branch, even if the Seidov limit is exceeded~\cite{Mendes:2024hbn}.  In order to counteract such errors, we do not use the Seidov limit to distinguish between absent and connected cases. Instead, we employ an entropy-dependent hybrid mass difference between $0.01\,\msol$ and $0.02\,\msol$ as the deciding factor. EoSs with smaller mass differences are considered to be absent of hybrid stars, even if numerically some hybrid stars might be located at the top of the first branch. 
    \\\\
    The presence of a second branch (i.e., cases B and D) provides a unique opportunity, because it allows two stars to have the same mass but vastly different radii. One of these stars has a quark matter core, while the other is purely hadronic. Such pairs are called twin stars; they are considered to be 'smoking gun' signals for phase transitions in neutron stars. Christian \etal~\cite{Christian:2017jni,Christian:2023hez} suggested a categorization scheme based on the maximum masses of both branches for these twin stars.\\
    Category I contains mass-radius relations in which both branches exceed $2\,\msol$, in category II only the first branch reaches $2\,\msol$. In category III only hybrid stars can reach $2\,\msol$ and category IV is a subset of category III, where purely hadronic neutron stars do not exceed $1\,\msol$, which would imply all currently known neutron stars to be hybrids. In Fig. \ref{fig:Christian_Cats} four representative examples for these categories are shown.\\  
    The goal of this categorization is to provide an intuitive guideline to infer phase transition properties from the visual changes of the mass-radius relation. Particularly, the distance between the first and second branch can be related to the strength of the phase transition, where larger values of $\Delta\epsilon$ increase the separation, while the shape of the second branch indicates the transition pressure (and by implication, the transition density). Higher values of $p_{trans}$ flatten the second branch and increase the length of the first one. This means categories I and II require a large transition pressure and feature second branches nearly parallel to the x-axis, while categories III and IV have much steeper second branches.

    \section{Purely Hadronic Proto-Neutron Stars}
    \label{sec:results_without_qm}
    \subsection{Entropy Dependence of the Maximum Mass Configuration}
     \label{sec:mb_evolution_without_qm}   
\begin{figure*}
        \centering
        \includegraphics[width=0.9\textwidth]{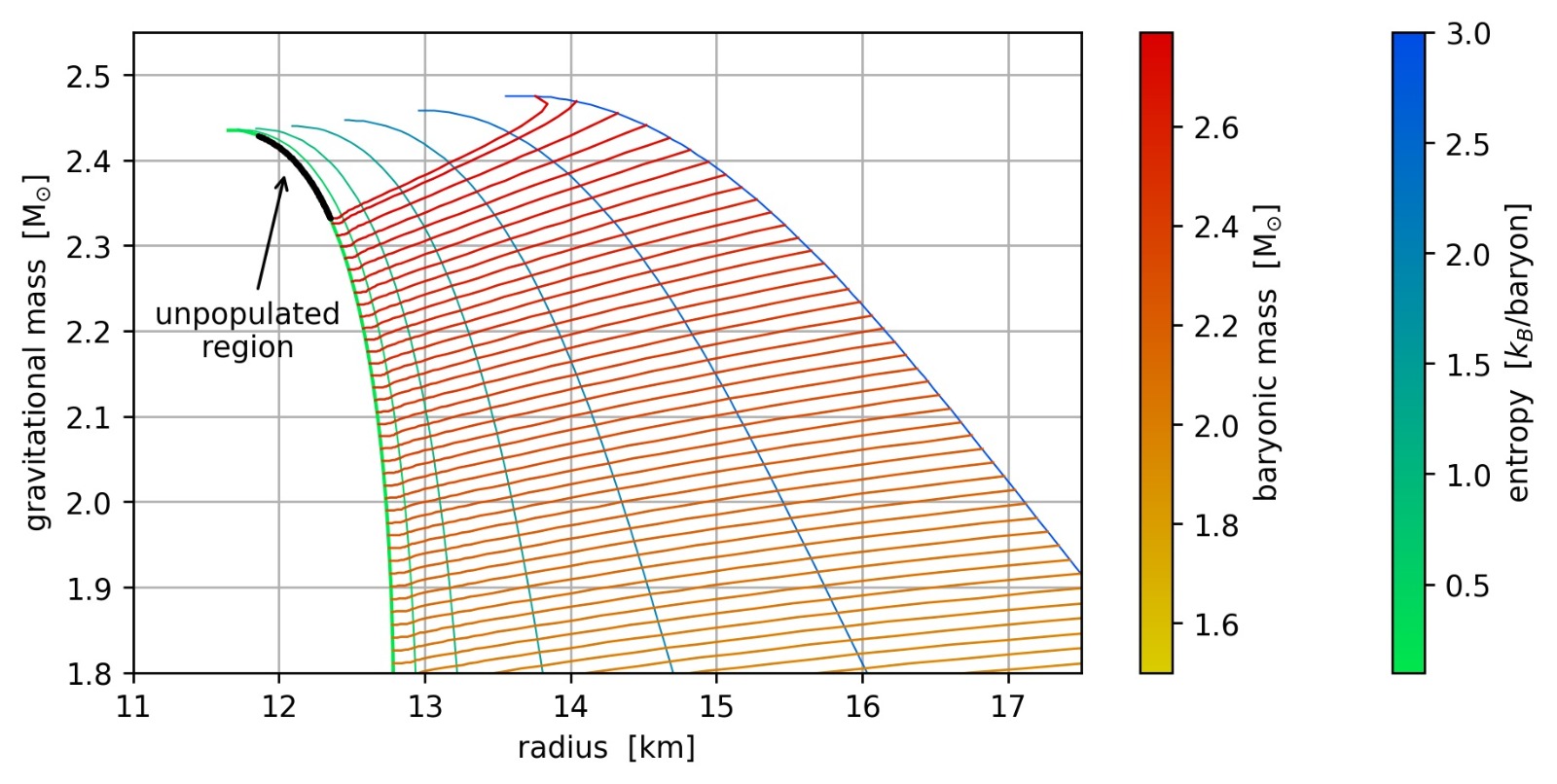}
        \caption{MR relations and evolutionary tracks of the FB65 EoS at $\yq=0.2$ with neutrinos. MR relations of different entropies ($\s=0.1$ and $\s=0.5$ to $\s=3.0$ in steps of 0.5) are shown in green and blue. Evolutionary tracks of PNSs with constant baryonic masses are shown in yellow and red. Only tracks with an initial entropy of $\szero=3.0$ are shown. The region marked in black on the lowest entropy MR relation contains stable neutron stars, but is not reached by any PNS tracks.}
        \label{fig:traces_fb65}
    \end{figure*}

\begin{figure*}
        \centering
        \includegraphics[width=\textwidth]{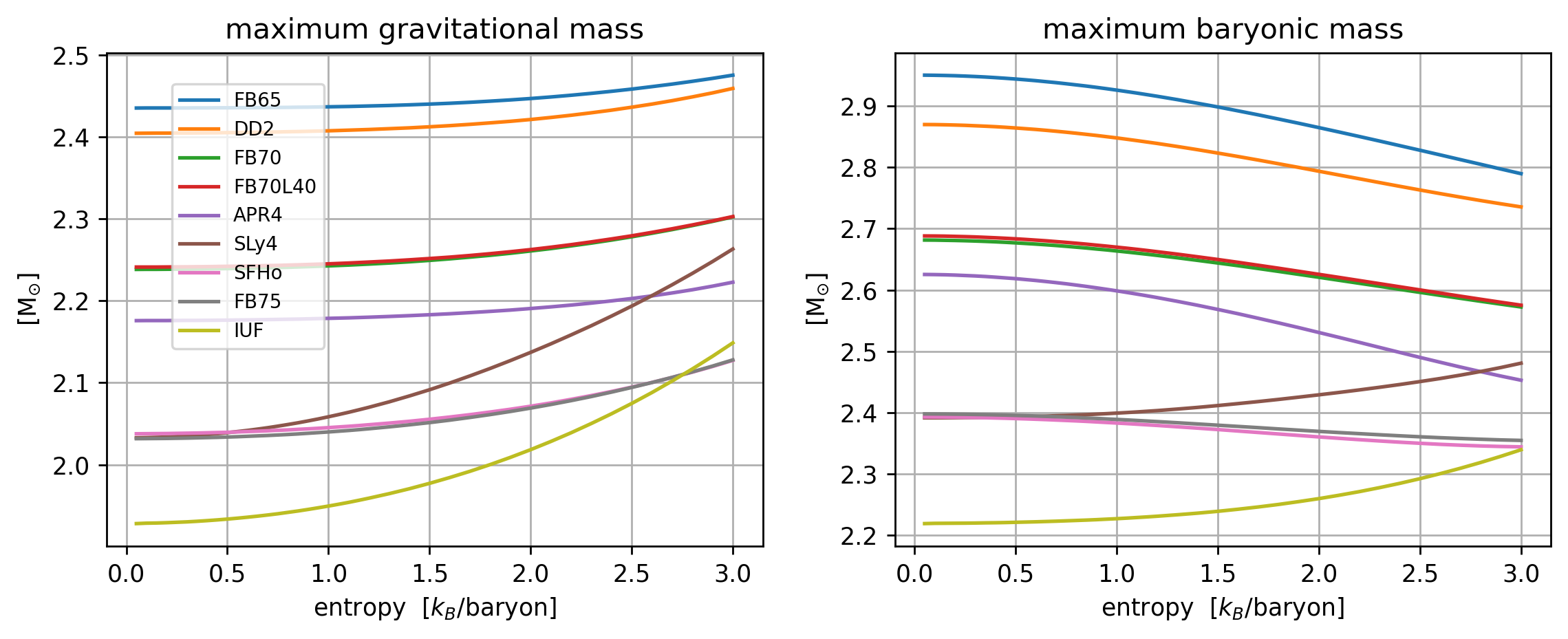}
        \caption{Maximum gravitational (left) and baryonic (right) masses as functions of entropy. All EoSs are compared at $\yq=0.2$. All maximum gravitational masses decrease with decreasing entropy. The maximum baryonic masses decrease with decreasing entropy for the SLy4 and IUF EoSs and increase for all other used EoSs. Note that the deviating behavior of SLy4 and IUF is caused by the flawed assumption of a constant charge fraction for the entire entropy range.}
        \label{fig:maximum_masses}
    \end{figure*}

    In this section we analyze the behavior of the maximal mass configuration of purely hadronic PNSs. 
    We find that all zero-temperature mass-radius relations have a maximum located at smaller (gravitational) masses than their high entropy counterparts. This is expected, as the binding energy of individual PNSs increases during the cooling process, therefore reducing their gravitational mass at constant baryonic mass. When comparing the baryonic mass-radius relation at low and high entropy stages, the opposite behavior emerges. The maximum (baryonic) mass is increasing as the star cools down. As a result, there is a region at the top of the cooled-down mass-radius relation with higher baryonic masses than the maximum mass at the initial entropy stage. Stars in that region are unstable during large parts of the cooling process and can therefore not be produced by CCSN. However, once a neutron star has cooled down, subsequent mass-accretion can populate this region. The mass range of this unpopulated region is highly dependent on the underlying EoS and on the initial entropy.
	An example for these unpopulated regions is shown in Fig. \ref{fig:traces_fb65} for the FB65 EoS. In this figure mass-radius relations for entropies between $\s=3.0$ and $\s=0.1$ are drawn. We track the evolution of individual PNSs along these mass-radius relations with lines of constant baryonic mass. The neutron stars at $\s=0.1$ that cannot be realized by CCSNe following our argument above are indicated by a thick black line.\\
    In Fig. \ref{fig:maximum_masses} we show the behavior of the maximum mass as a function of entropy for all the EoSs we used in this work. On the left side the maximum gravitational mass is shown, and on the right side the baryonic mass. There are two EoSs for which the maximal baryonic mass decreases with entropy, SLy4 and IUF. This would imply a destabilization of very massive PNS during the cooling process, instead of a region of unpopulated masses. However, we stress that this is a side effect of assuming a constant charge fraction throughout the PNS evolution. When considering two ``snapshots'' to approximate a more realistic scenario at a high charge fraction for $\s=3.0$ and beta-equilibrium at $\s=0.1$, these two EoSs also feature an unpopulated region. Likewise, the unpopulated region of the remaining EoSs increases for such a comparison.\\
    All EoSs used by us are without strangeness. However, if hyperons are included in the EoS, their onset leads to a softening of the EoS, which, in turn, can lead to a decreasing maximum baryonic mass during the cooling phase, even under more realistic conditions~\cite{Glendenning:1994za,Keil:1995hw,Bombaci:1996,Prakash:1997,Pons:1999,Prakash:2001}.

    \subsection{Proto-Neutron Star Density Profiles}
    \label{sec:structure_evolution}
\begin{figure*}
        \centering
        \includegraphics[width=\textwidth]{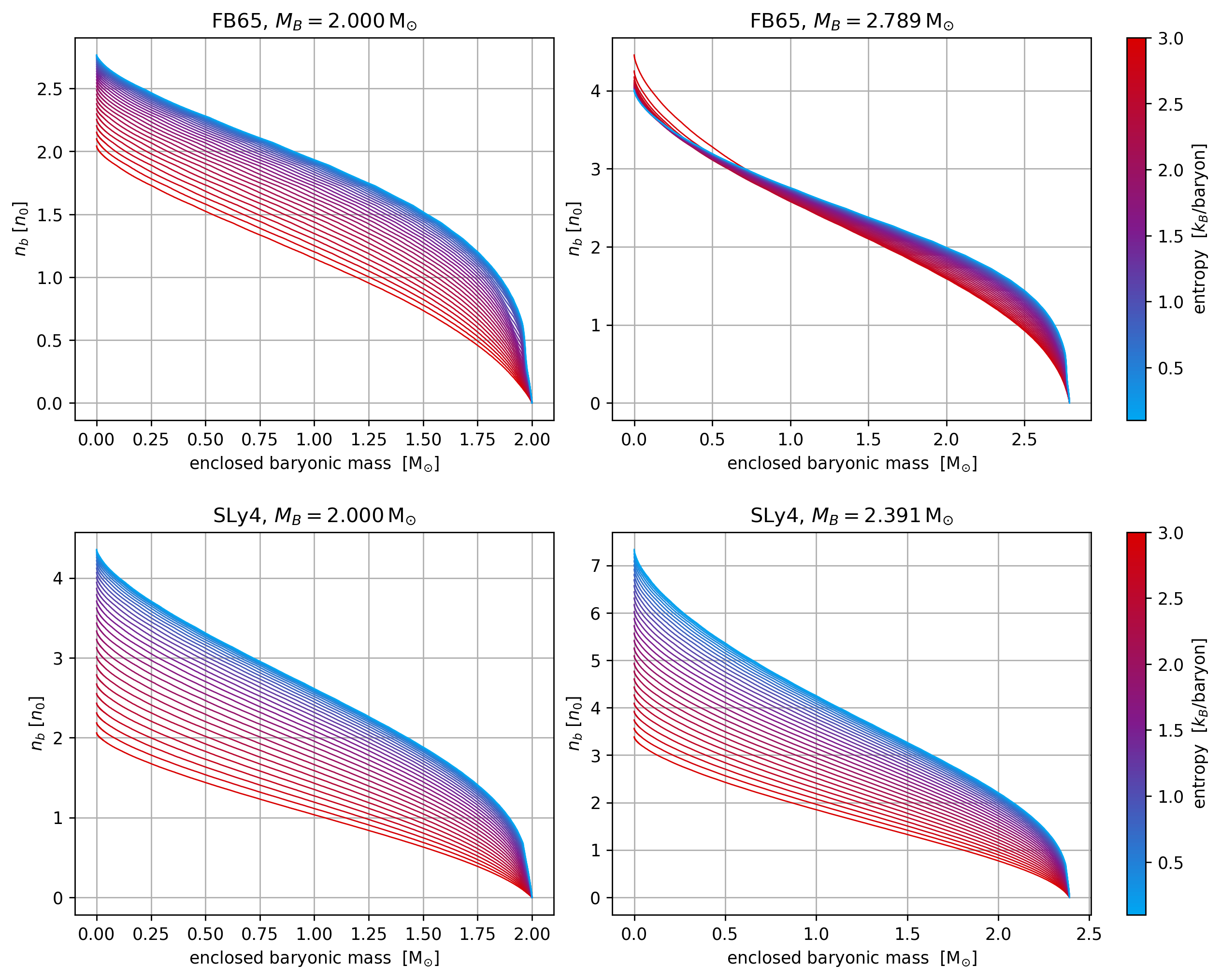}
        \caption{Radial profiles of the baryon density $\nb$ (in units of the nuclear saturation density $\nzero$) for all considered entropies from $\s=3.0$ (red) to $\s=0.1$ (blue) in steps of 0.1, compared for different EoSs and baryonic masses. Because the radii change during the evolution, the enclosed baryonic mass is used as a Lagrangian x-axis coordinate. The used EoSs are FB65 (top panels) and SLy4 (bottom panels) at $\yq=0.2$ with neutrinos. The left panels show the profiles of PNSs with a baryonic mass of $2.0\,\msol$, corresponding to cold gravitational masses of $1.765\,\msol$ and $1.759\,\msol$. The right panels show profiles of PNSs with the maximum baryonic mass that is stable for all entropies $0.1 \leq \s \leq 3.0$. This results in baryonic masses of $2.789\,\msol$ and $2.391\,\msol$, corresponding to cold gravitational masses of $2.331\,\msol$ and $2.033\,\msol$. The entropies/colours can be thought of as proxies for the temporal evolution (``red= early, blue= late'').}
        \label{fig:density_profiles}
    \end{figure*}

    Here we explore the density profiles during the cooling process of PNSs. 
  	In general, a hot PNS features a larger radius than its cooled-down counterpart with the same baryonic mass. 
    This means globally the density of the star is increasing when cooling down. 
    For most stars, this is also locally true, i.e., for every fixed enclosed baryonic mass, the density is increasing as the entropy is decreasing. In other words, the density profile is shifted to higher densities as the PNS cools down, but it retains its general shape, the hot core and hot mantle are less dense than the cold core and crust, respectively. This is not the case for the most massive neutron stars. Counterintuitively, their core region becomes less dense as the star is cooling down, while the outer regions behave normally. Nevertheless, even for the most massive stars, the density rises monotonically from the PNS surface to the core.
    We give an example of this behavior in the first row of Fig. \ref{fig:density_profiles}, where the density profile at various entropy stages is shown. We use the FB65 EoS, which is a particularly stiff case and therefore features a large maximum mass. On the left side, the density profiles of a $M_{B} = 2\,\msol$ PNS are shown, on the right side, the maximum stable mass. For the $M_{B} = 2\,\msol$ case, the density profile retains its general shape during the entire evolution; only the drop in density at the PNS surface becomes more pronounced as the star cools down. This is the expected behavior that most stars exhibit. However, on the right side of Fig. \ref{fig:density_profiles} significant changes in the density profile can be seen during the cooling process. As the star cools down, the core decreases in density, while the outer regions increase in density.
     \\\\
   This unexpected behavior warrants a closer look. The most massive track of Fig.~\ref{fig:traces_fb65} corresponds to the density profile shown in the upper right panel of Fig. \ref{fig:density_profiles}. Since its gravitational mass is monotonically decreasing, the binding energy of the entire PNS is increasing, even as the core becomes less dense. Therefore, we can assume that these evolutionary paths are energetically stable. Furthermore, we verified that this behavior is not caused by our assumption of flat charge and entropy profiles by again comparing two density snapshots of a high charge fraction at $\s=3.0$ and beta-equilibrium at $\s=0.1$. Nevertheless, we cannot fully rule out that this behavior is due to our quasi-static approach and it might not occur in a fully dynamical simulation.
   \\\\
   If a PNS features such an unusual density profile evolution, it implies an expanding core region and a shrinking mantle as the star cools down. As a result, it is possible for particularly massive PNSs to increase their total radius during the early cooling stages if the initial entropy was sufficiently high. An example of this can be seen in Fig.~\ref{fig:traces_fb65}, where the radius of the most massive track increases slightly for the first entropy stages before the PNS starts to contract normally.
   \\\\
   The extent to which this counterintuitive density profile is exhibited depends on the EoS. Some EoSs do not show this behavior at all. An example of an EoS where the density profile shows the same behavior for all its masses is depicted in the second row of Fig. \ref{fig:density_profiles} for SLy4. It should be noted that the maximum mass of SLy4 is significantly smaller than FB65. For the EoSs that we investigated, we found that FB65, FB70, DD2, and APR4 show a decreasing central density at high masses and entropies. For the FB75 and SFHo EoSs, the shape of the density profile changes significantly, but the central density at high entropies remains roughly constant. The SLy4 and IUF EoSs show qualitatively similar density profiles for all stable masses.

    \section{Proto-Neutron Stars with Phase Transition}
    \label{sec:results_with_qm}
    \subsection{Entropy Dependence of the Maximum Mass Configuration}
    \label{sec:pns_evolution_with_qm}
\begin{figure*}
        \centering
        \includegraphics[width=\textwidth]{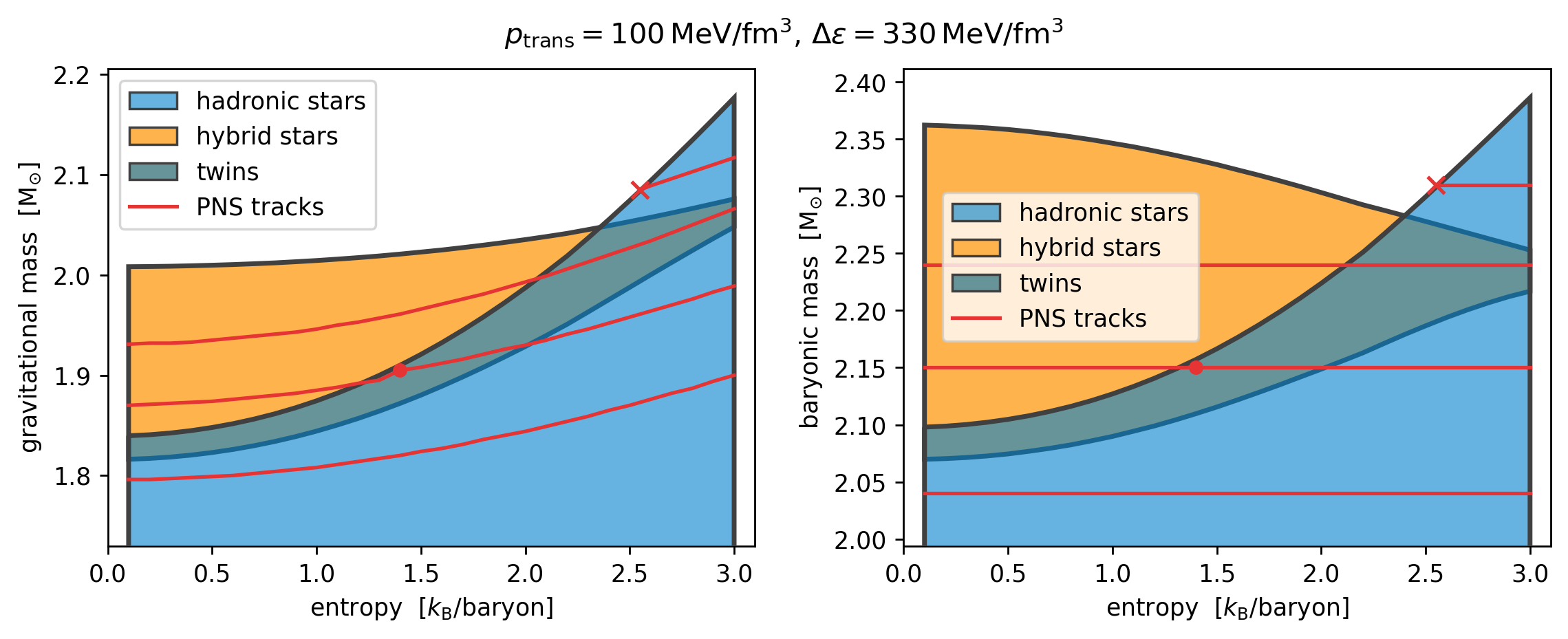}
        \caption{Mass ranges of the hadronic and second hybrid stars (blue and orange) as functions of entropy. Both areas are plotted partially transparent so that the overlapping area is visible. The overlapping area contains twin stars. The used CSS configuration has $\ptrans = 100\,\mevperfmcub$, $\Delta\epsilon = 330\,\mevperfmcub$ and $\cqm = 1$. The left panel shows gravitational masses, the right panel baryonic masses. The evolution of PNSs with a constant baryonic mass follows horizontal lines in the right panel. Four exemplary PNS evolutions with baryonic masses of $2.04\,\msol$, $2.15\,\msol$, $2.24\,\msol$ and $2.31\,\msol$ are shown with red lines. The PNS with $\mbary = 2.04\,\msol$ (bottom line) remains purely hadronic. The PNS with $\mbary = 2.15\,\msol$ (second line from the bottom) reaches the transition pressure $\ptrans(\s)$ at an entropy around $\s \approx 1.4$, marked by a red dot. The PNS with $\mbary = 2.24\,\msol$ (second line from the top) can be a hybrid star for the whole range of entropies. The PNS with $\mbary = 2.31\,\msol$ (top line) becomes unstable at an entropy $2.5 < \s < 2.6$, marked by a red X.}
        \label{fig:mrange_evolution_demo}
    \end{figure*}

\begin{figure*}
    	\centering
    	\includegraphics[width=0.90\textwidth]{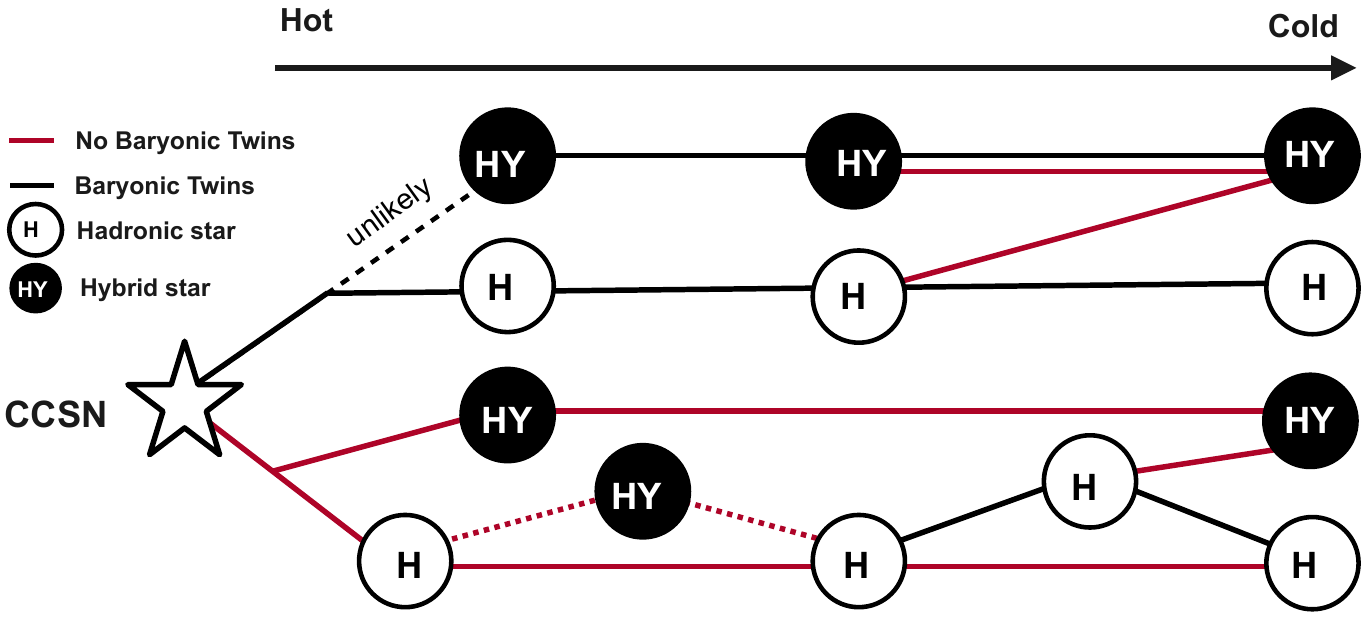}
    	\caption{Schematic depiction of all the evolutionary paths available to PNSs with a fixed baryonic mass. White circles stand for a purely hadronic PNS and black circles for a hybrid PNS. Black lines indicate that the given mass of the PNS supports a hadronic and a hybrid solution, i.e. baryonic twins are possible. The dashed black lines shows that, according to CCSNe simulations, it is unlikely for a hybrid PNS to form, if a hadronic PNS of the same mass is also possible~\cite{Naseri:2024}. Baryonic twins can appear and disappear at different stages of cooling process, which is highly dependent on the underlying EoS and phase transition parameters. Red lines indicate that there is either a stable hadronic or a stable hybrid star, but not both, i.e. there are no baryonic twins. The dashed red lines show the rare case, where we found a quark core that forms only temporary and vanishes at smaller entropy, see also Kunkel \etal~\cite{Kunkel:2026wvz}, who outline a similar scenario for a more involved quark model.\\
       }
    	\label{fig:Flowchart}
    \end{figure*}

  In this section, we analyze the changes to the PNS evolution brought on by a strong phase transition. 
  Like in the previous section \ref{sec:results_without_qm} the maximum (baryonic) mass configuration increases during the PNS cooling evolution, causing the top of the mass-radius relation to be unpopulated. The introduction of a phase transition does not alter this behavior, however, its presence means that the most massive neutron stars usually are hybrid stars. 
  Perhaps surprisingly, we find that the maximum baryonic mass of purely hadronic stars changes its behavior compared to a setup without a phase transition. Its maximum mass decreases as the star is cooling down. This counterintuitive behavior can be attributed to the fact that the entropy-dependent transition point determines the end of the hadronic part, instead of maximum mass itself. This transition point, given by $\ptrans(s)$, decreases with decreasing entropy, while the central pressure of a PNS usually increases during the cooling process.
  The opposite behavior of the maximum baryonic masses of hybrid and hadronic stars was also described by Carlomagno \etal~\cite{Carlomagno:2024} for the DD2 EoS with a NJL quark phase. 
  \\\\
Based on this opposite behavior, we find four broad scenarios for the evolution of PNSs in our model.  
\begin{enumerate}
   		\item The first scenario is the simple case where a PNS that does not contain quark matter in its initial stages and does not reach a central pressure higher than $\ptrans(s)$ will remain hadronic throughout its evolution.
   		\item PNSs of the second scenario are initially purely hadronic, but at some stage during their evolution no stable hadronic star with their baryonic mass is possible. If a stable hybrid star with the same baryonic star is available instead, the PNS will contract until it reaches equilibrium again. The resulting neutron star will be a hybrid star. 
   		\item The third possible evolution for a PNSs starts as a hybrid star. This would imply that the transition already occurs during the CCSN. All initially hybrid PNSs in our setup remain hybrid stars during their evolution. 
   		\item Particularly massive, purely hadronic PNSs follow the fourth scenario. These PNSs become unstable during their evolution without a stable hybrid star configuration with the same baryonic mass being available. This is a direct result of the maximum mass of purely hadronic stars decreasing if the EoS contains a phase transition. 
   \end{enumerate}
    Exemplary tracks for these scenarios are shown in Fig.~\ref{fig:mrange_evolution_demo} as red lines. The figure shows the gravitational and baryonic mass ranges of the hadronic and hybrid stars as functions of entropy in blue and orange, respectively. The EoS is parameterized by $\ptrans = 100\,\mevperfmcub$ and $\Delta\epsilon = 330\,\mevperfmcub$. 
  This configuration can be classified as category III following Christian \etal~\cite{Christian:2017jni} and has a disconnected hybrid branch for the entire entropy range. 
  Where the two mass ranges overlap, twin stars are present. Note the opposite entropy dependence of hadronic and hybrid maximal baryonic masses displayed on the right side of the figure.
\\\\
In Fig.~\ref{fig:Flowchart} we show a schematic depiction of all evolutionary paths a given PNS can take. Every path falls into one of the four stated scenarios. The exact path through this diagram is determined by the (baryonic) mass of the PNS and the underlying equation of state. Furthermore, we only show paths that end in a stable solution. A key concept in the behavior of a PNS is the presence of a so-called baryonic twin, which we will discuss in the next section.

    \subsection{Twins in the Shadow}
    \label{sec:twins_in_the_shadow}
        \begin{figure*}
    	\centering
    	\includegraphics[width=\textwidth]{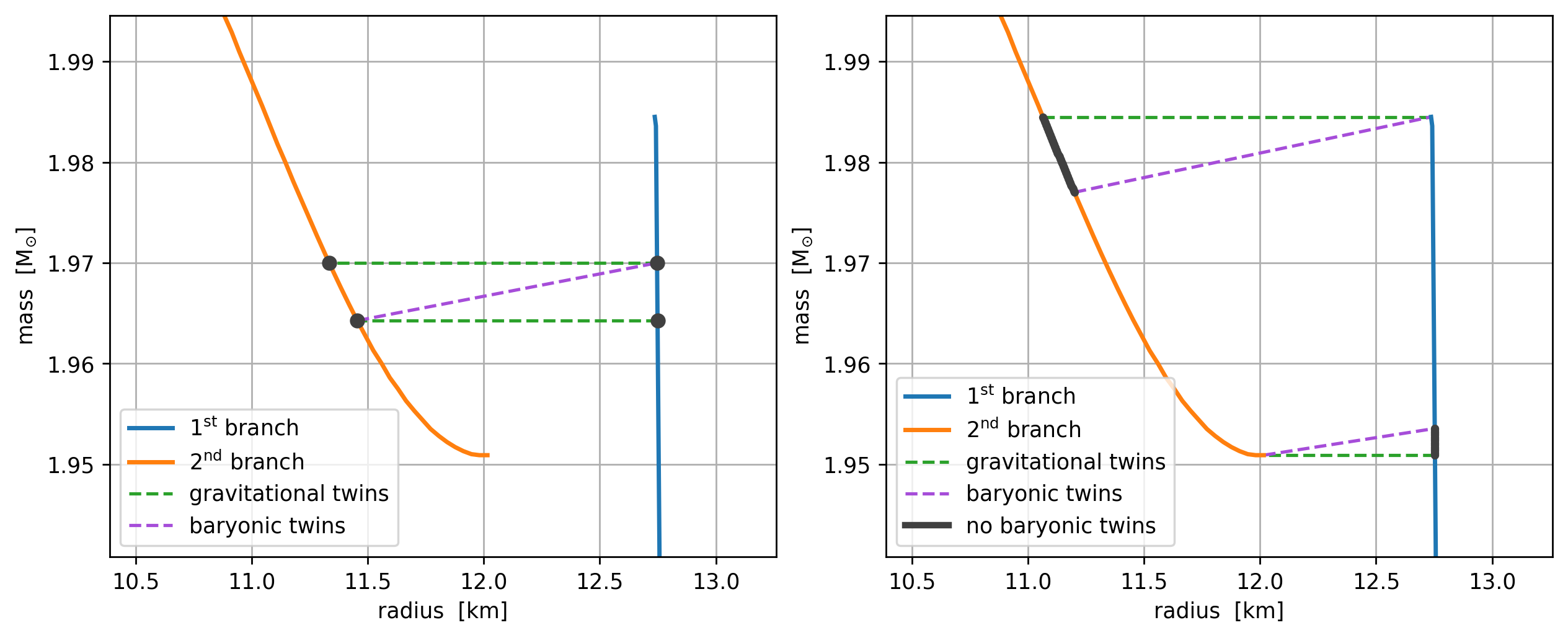}
    	\caption{Examples of gravitational and baryonic twins in the cold ($\s = 0.1$) MR relation with $\ptrans = 126\,\mevperfmcub$ and $\Delta\epsilon = 340\,\mevperfmcub$. Pairs of gravitational and baryonic twins are connected with green and purple dashed lines. The left panel shows one exemplary pair of baryonic twins and their corresponding gravitational twins. The right panel shows both types of twins of the maximum mass first branch star and the minimum mass second branch star. Parts of the MR relation that have gravitational twins, but no baryonic twins, are marked with a gray line.}
    	\label{fig:baryonic_twins}
    \end{figure*}
\begin{figure*}
        \centering
        \includegraphics[width=0.7\textwidth]{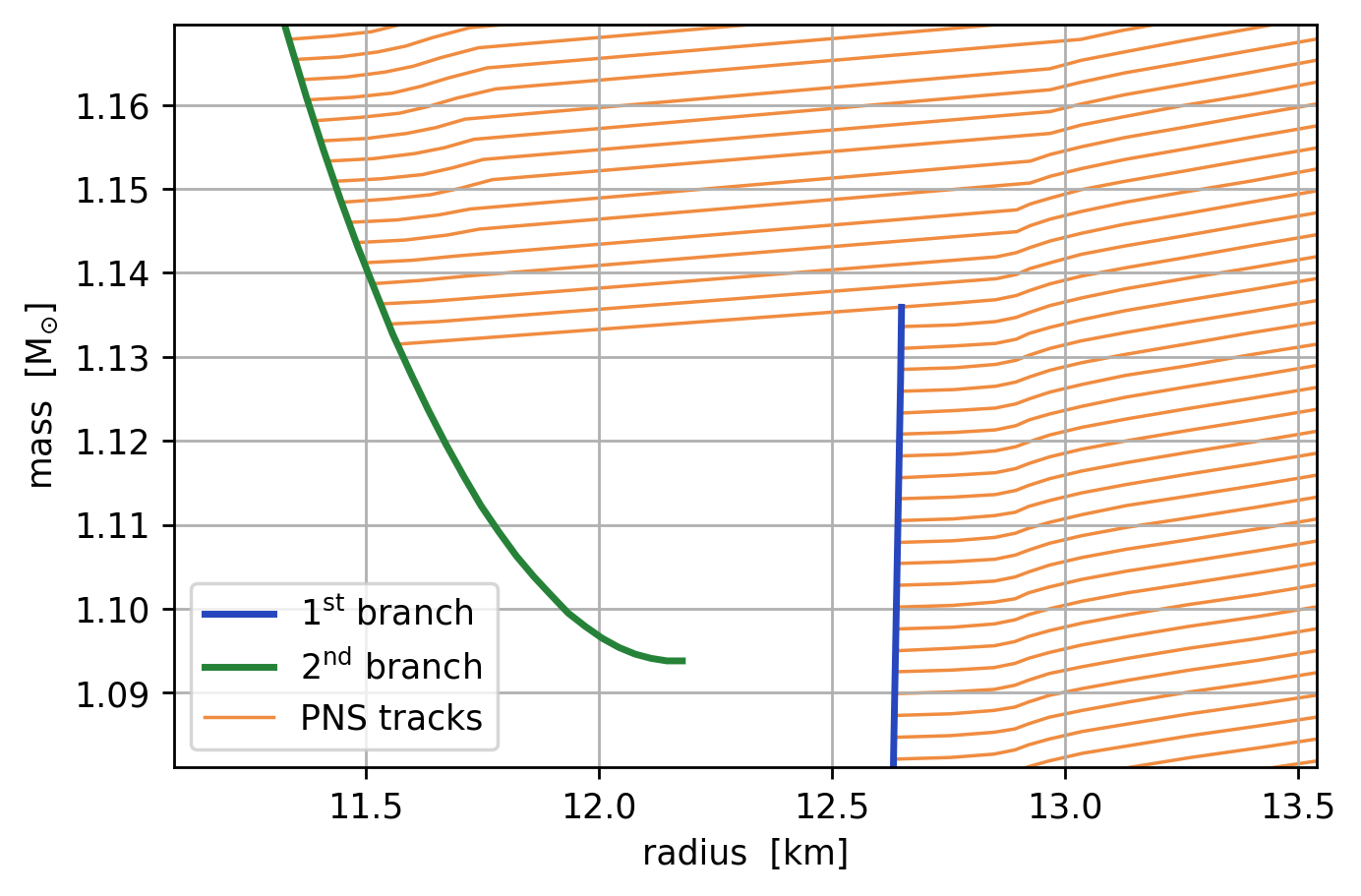}
        \caption{MR relation with $\ptrans = 36\,\mevperfmcub$ and $\Delta\epsilon = 340\,\mevperfmcub$, zoomed in to the twin star region. Evolutionary tracks of PNSs with different baryonic masses are shown as orange lines. The second branch baryonic twins are not reached by any tracks because they lie in the shadow of the first branch.}
        \label{fig:twins_in_the_shadow}
    \end{figure*}

    In this section we examine the special case of twin stars in more detail. For this purpose, we need to differentiate between baryonic and gravitational twins. We find that hybrid stars are generally stronger bound and therefore have a smaller gravitational mass. An example of this is shown on the left side of Fig.~\ref{fig:baryonic_twins} with a purple dashed line.\\
    On the right side of the same figure, the baryonic and gravitational twins of the first branch maximum and the second branch minimum are shown. Baryonic twins are again indicated by purple dashed lines, and gravitational twins are shown by green dashed lines. Due to the difference in binding energy, there are regions where gravitational twins do not have a baryonic twin. Specifically, these regions are located at the upper end of the twin star mass range on the hybrid star branch and at the lower end of the twin star mass range on the hadronic branch. All other neutron stars in the twin star mass range have a baryonic twin.
    This is important for the evolution of PNSs, because baryonic twins must have followed different evolutionary paths despite having the same baryonic mass, making it likely that only one of them is realized in nature.
    \\\\
    Under our assumption of equilibrium, no hadronic PNS reaches the necessary central pressure to become a hybrid PNS, as long as there is also a stable hadronic solution available. The hydrodynamical simulations performed by Espino \etal~\cite{Espino:2022} support this point. They perturbed configurations in the unstable region between the hadronic and hybrid branches, leading to radial oscillations and bouncing between both phases. Ultimately, these configurations settled onto the stable hadronic branch. This suggests that a baryonic hybrid twin may need to be produced in a quasistatic way to circumvent the bounce back (see also \cite{Naseri:2026xrq}). 
    However, this is not possible in the model we investigate here, since reaching the phase transition point would trigger a collapse on the dynamical time scale. 
    Therefore, a hadronic PNS cannot become a hybrid PNS unless it is the only stable solution. In other words, we find that no cooled-down hybrid star can have a baryonic twin, with only one exception: the initial PNS is already a hybrid star with a baryonic twin. This can be checked by comparing Fig.~\ref{fig:Flowchart}, which shows all PNS paths in our model.
    \\\\
    However, CCSN simulations performed by Naseri \etal~\cite{Naseri:2024} suggest that only the hadronic solution can be realized, even if both a hadronic and a hybrid PNS are possible for a given baryonic mass. The collapsing core first reaches the stable hadronic solution, and the collapse stops there. The core bounce happens below the transition pressure even for extreme initial perturbations. Nevertheless, Naseri \etal~\cite{Naseri:2024} find that once the PNS starts to cool and the transition point is reached, the resulting collapse can be stabilized before the PNS turns into a black hole.
    \\\\
    Following this line of reasoning, we find that hybrid stars can only be formed if they have no baryonic twin. Fig.~\ref{fig:twins_in_the_shadow} highlights this by showing a representative mass-radius relation with evolutionary tracks of different PNSs. The area not reached by any track on the second branch appears to be ``in the shadow'' of the first branch. We therefore refer to this phenomenon of unpopulated twin stars as ``twins in the shadow''.

    \section{Parameter Space Analysis}
    \label{sec:css_categorization}
    \begin{figure*}
    	\centering
    	\includegraphics[width=0.8\textwidth]{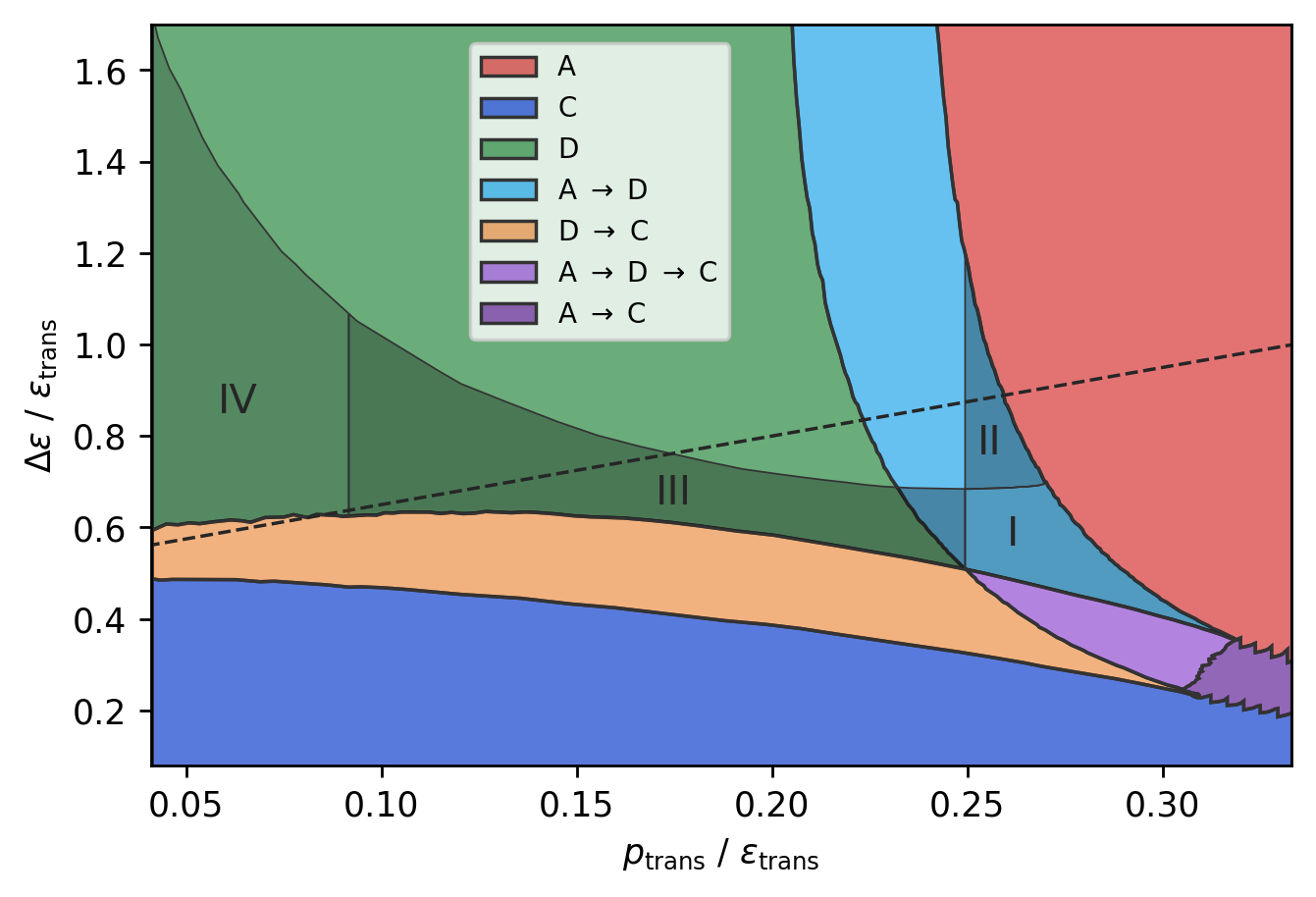}
    	\caption{Parameter space of hybrid branch topologies considering finite entropies. Shown are seven distinct categories: Connected (C, dark blue), Disconnected (D, green), Absent (A, red), Disconnected at high entropy and Connected at low entropy ($\catdc$, orange), Absent at high entropy and Disconnected at low entropy ($\catad$, light blue), temporarily disconnected ($\catadc$, light purple), and direct transition from Absent to Connected ($\catac$, dark purple). Categories D and $\catad$ contain twin stars at the lowest entropy. The twin star categories $\catitoiv$ by Christian \etal~\cite{Christian:2017jni} are shown as dark shaded areas on top of the other categories. The Seidov limit is shown as a dashed line.}
    	\label{fig:hbb_parameter_space}
    \end{figure*}
      \begin{figure*}
    	\centering
    	\includegraphics[width=\textwidth]{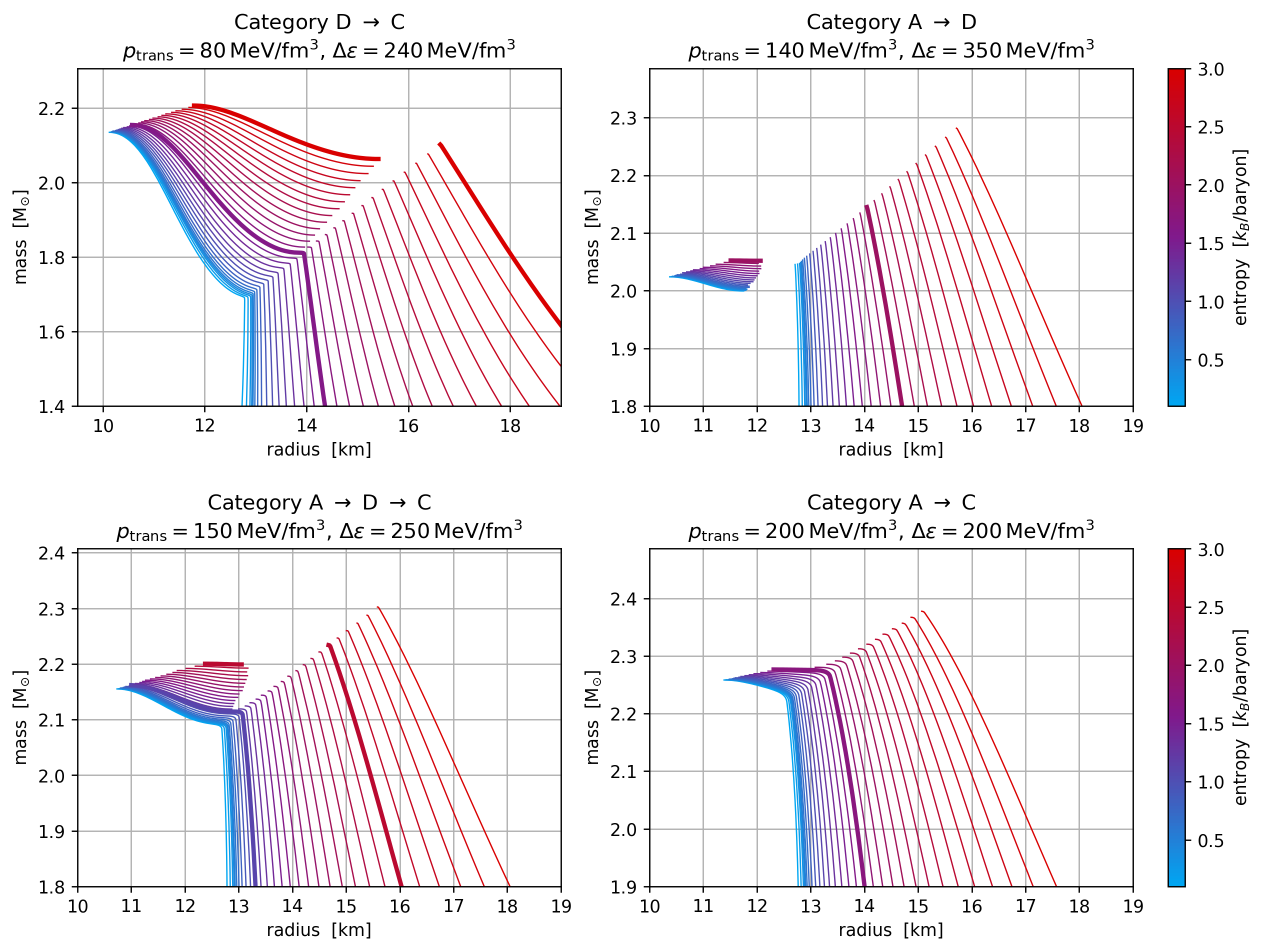}
    	\caption{Each panel shows MR relations for varying entropies between $\s = 0.1$ (blue) and $\s = 3.0$ (red) for the FB65 EoS with a CSS quark phase with $\cqm = 1$. In the top left panel (category $\catdc$), the hybrid branch is disconnected at high entropies and connected at low entropies. In the top right panel (category $\catad$), the hybrid branch is absent at high entropies and disconnected at low entropies. In the bottom left panel (category $\catadc$), the hybrid branch is absent at high entropies, connected at low entropies, and disconnected at intermediate entropies. In the bottom right panel (category $\catac$), the hybrid branch is absent at high entropies and connected at low entropies. The highest entropy MR relations that have a disconnected or connected hybrid branch are plotted with thick lines to make it easier to distinguish between the categories.}
    	\label{fig:mr_evolution_hbb_2}
    \end{figure*}
\begin{figure*}
    	\centering
    	\includegraphics[width=\textwidth]{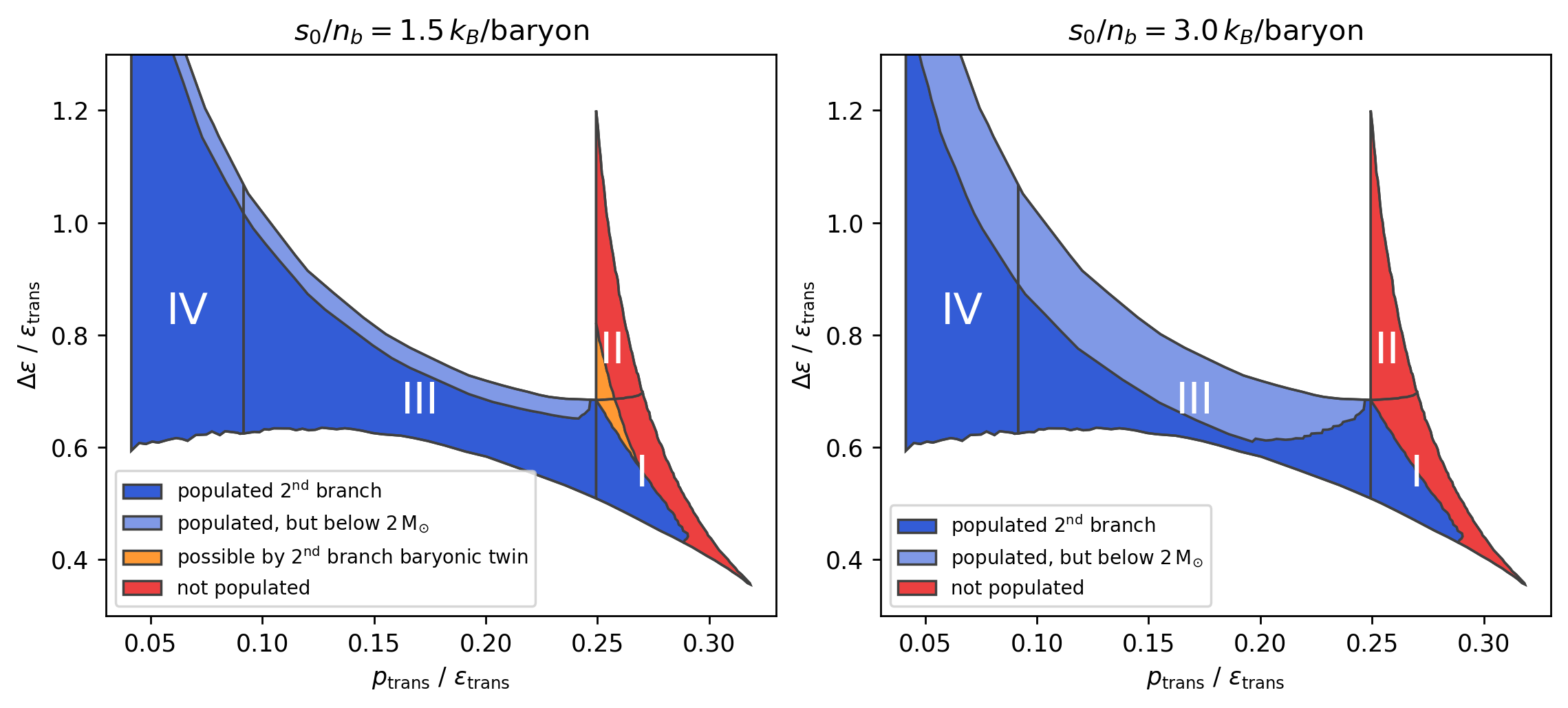}
    	\caption{CSS parameter space of the four twin star categories, colored by whether the second branch can be populated. The left panel uses an initial entropy of $\s=1.5$, and the right panel an initial entropy of $\s=3.0$. In the blue-colored regions, the second branch can be reached by cooling PNSs. In the light blue regions, the maximum populated mass lies below $2\,\msol$. In the red-colored regions, no part of the second branch can be reached by cooling PNSs. In the orange-colored region, which is only present for initial entropies of $\s \lesssim 2.0$, the second branch could be populated by a PNS that is a hybrid baryonic twin at the initial entropy. It is unlikely that such a PNS could be formed, but this is outside the scope of this work.}
    	\label{fig:second_branch_population}
    \end{figure*}

    In this section we investigate the full parameter space of the CSS parameterization. To this end, we visualize the $\Delta\epsilon/\epstrans$-$\ptrans/\epstrans$ space in Fig. \ref{fig:hbb_parameter_space} and highlight the different categories of hybrid stars. The division by $\epstrans$ is well known to allow for easier comparison with other EoSs.\\
     We find that it is still possible to classify all mass-radius relations with the ABCD scheme introduced by Alford \etal~~\cite{Alford:2013aca} (see section~\ref{sec:hybrids_and_twins}). At different entropy stages, however, one EoS can fall into different categories. Naturally, this is particularly the case in the parameter ranges where the categories border each other. If hadronic and hybrid branch are separated, the jump in energy density governs the extent of separation, as outlined by Christian \etal~\cite{Christian:2017jni}. The impact of $\Delta\epsilon$ on the mass-radius relation appears to be heightened during the early stages of the PNS cooldown.\\
     As a result, four new intermediate regions appear in Fig. \ref{fig:hbb_parameter_space}. For example, a mass-radius relation might feature a detached second branch at high entropies that connects later on, it undergoes a D $\rightarrow$ C evolution. This happens at small values of $\Delta\epsilon$, i.e., the region where cases C and D meet. Similarly, evolutions  A $\rightarrow$ D, A $\rightarrow$ C, and even A $\rightarrow$ D $\rightarrow$ C can be realized. Note that all category I and II twin stars are located in the A $\rightarrow$ D region. Examples for these four scenarios are shown in Fig. \ref{fig:mr_evolution_hbb_2}. The mass-radius relations where the classification changes are indicated by a thicker line.
     \\\\
     While not shown explicitly in Fig. \ref{fig:mr_evolution_hbb_2}, the shape of the A $\rightarrow$ D case heavily suggests that parts of the hybrid branch, if not the entire branch, is in the shadow of the first and can therefore not be formed from PNS. We confirm this suspicion in Fig. \ref{fig:second_branch_population}, where we show the parameter space in which twin stars are possible, see section \ref{sec:twins_in_the_shadow}.
     The blue-colored regions indicate parameter sets where the hybrid star branch can be populated. It contains a light blue part, where a second branch exists, but $M_{\mathrm{star}}\ge2\,\msol$ solutions cannot be populated due to an insufficiently massive stable PNS (see section \ref{sec:mb_evolution_without_qm}). These parameter sets are not entirely ruled out by the well-established mass measurements since later mass accretion is possible, but they cannot meet astrophysical constraints immediately after the PNS evolution.
     In the red-colored regions, no part of the second branch can be reached by cooling PNSs because the second branch is located completely in the shadow of the first branch. Finally, the orange region corresponds to parameter sets where the only stable hybrid stars at the initial entropy have baryonic, purely hadronic twins, which likely means that they are not realized in nature (see section \ref{sec:twins_in_the_shadow}). Since the second branch in the red and orange area is located entirely at baryonic masses smaller than the maximum baryonic mass of the first branch, it cannot be populated through accretion once cooled down.
	 \\\\
     On the left side of Fig. \ref{fig:second_branch_population} we assume an initial entropy of $\s=1.5$. This represents our most generous scenario, since most CCSN simulations predict larger values~\cite{Pons:1999,Prakash:1997,Pascal:2022,Roberts:2012,Nagakura:2020,Jakobus:2022,Jakobus:2025}. In this scenario most EoSs support $2\,\msol$ stars immediately after cooling down. Furthermore, the second branches of categories III and IV can be populated in their entire parameter space. Category I can be partially populated, and category II is completely ruled out, unless the EoS supports baryonic twins during the entire cooling process.\\
     On the right side of Fig. \ref{fig:second_branch_population} we increase the initial entropy to $\s=3.0$. This significantly increases the area where $2\,\msol$ cannot be populated in categories III and IV. However, these two categories remain otherwise unaffected by the increased initial entropy. Contrarily, category II is completely ruled out by the increase in initial entropy. Especially for categories I and II an entropy of $\s=3.0$ is a more likely scenario, as it has been shown that the initial entropy of a PNS is proportional to the mass of its progenitor \cite{Temaj:2023nuo}.
    
	\section{Conclusion}\label{conclusion}

In this work, we systematically studied the evolution of proto-neutron (PNS) stars with and without phase transitions to quark matter. We approximated the cooling phase of PNSs by taking ``snapshots'' of the mass-radius relation at decreasing entropy stages. In addition, we assumed flat profiles of entropy and charge fraction for each PNS. By assuming that no mass is accreted during the cool-down, we were able to track the evolutionary path of individual PNSs by following constant baryonic masses through the mass-radius relations. 
    \\\\
    In section~\ref{sec:results_without_qm} we investigated purely hadronic cases and outlined that the maximum baryonic mass configuration is decreasing with entropy, in other words, the slope of $M_{B,max}(s)$ is negative. This leads to a range of masses at the top of a zero temperature mass-radius relation that cannot be populated by the PNS evolution alone. The size of this unpopulated region is determined by the EoS. 
    Furthermore, we found that for some EoSs the most massive stars exhibit a counterintuitive behavior in their density profile. While the average density is increasing as the star cools down, some EoSs show a decreasing central density compared to their hot counterparts. This counterintuitive behavior is not present for all EoSs, but seems to be associated with EoSs that feature high maximum masses.
    \\\\ 
    In section~\ref{sec:results_with_qm} we added a phase transition to one of the stiffer EoSs of the previous section, which gives us the largest parameter space to explore. We found that the top of the zero temperature mass-radius relation still has an unpopulated region at its top, even with the onset of quark matter. However, if the EoS contains a phase transition, the maximum baryonic mass of purely hadronic stars decreases as the star cools down, because it is determined by the transition pressure, which decreases with decreasing entropy. We identify four broad scenarios for a proto-neutron star's evolution:
    \begin{enumerate}
    \item The transition is not triggered at all, the PNS will remain purely hadronic.
    \item The transition takes place during the evolution, once the necessary pressure is reached in the center of the PNS. The PNS will become a hybrid star.
    \item The transition takes place during the supernova, i.e. the initial PNS is already a hybrid and will remain so.
    \item The PNS collapses or explodes before a stable hybrid star with the same baryonic mass can be formed. 
    \end{enumerate}
    We further find that a hadronic PNS will not reach a sufficient central pressure to trigger a phase transition, as long as a stable hadronic solution is possible. This affects the second scenario significantly, since a hadronic star can only become a hybrid star, if it stars to collapse. As a result we find that a cold hybrid star can only be produced, if it has no baryonic twin, i.e. a hadronic star with the same baryonic mass. The only exception to this rule would be the third scenario, where the initial PNS is already a hybrid. However, dynamical simulations (e.g. Naseri \etal\cite{Naseri:2024}) suggest that the hybrid twin is less likely to form. In an even more recent work Naseri and Paschalidis~\cite{Naseri:2026xrq} reinforce this point, by showing that a hadronic twin is favored, despite the hybrid twin being energetically favorable. They argue that this is due to the realistic cooling timescale being significantly longer than the timescale of a collapse. 
    We refer to the phenomenon that cooled down hybrid stars cannot have a baryonic twin as ``twins in the shadow'', since the stable hybrid star solutions on the disconnected branch of a mass-radius relation appear to be ``in the shadow'' of the purely hadronic branch. While baryonic twin stars are ruled out in this approach gravitational twin stars are still possible, because hadronic and hybrid stars have different binding energies.
    \\\\
    Finally, in section~\ref{sec:css_categorization} we show that this ``twins in the shadow'' phenomenon has a strong impact on the feasibility of twin stars, especially if the phase transition takes place at high pressures and features a large jump in energy density. In such cases the disconnected hybrid star branch can be fully within the shadow of the hadronic branch, ruling out twin stars for these parameter sets entirely. We also show that this can be somewhat mitigated by assuming relatively small initial entropy for the PNS, but CCSN simulations suggest that higher masses imply higher initial entropy~\cite{Temaj:2023nuo}.
    \\\\
    By employing the constant speed of sound parameterization for quark matter and a quasi-static approach for the PNS evolution we made our approach as general as possible. However, this also means that we had to make some approximations, which could effect our results. In particular, we could not investigate the timescale of the evolutionary steps, since in our approach there is no suitable way to approximate how much time passes between snapshots in our quasi-static approach. A full dynamical simulation of particularly interesting paths could help to better understand the underlying physical processes. Furthermore, our approach implicitly assumes a rapid conversion rate from hadronic and quark matter, but a slow conversion rate is hypothesized to stabilize parts of the region separating hadronic and hybrid stars. Changing the conversion rate could potentially increase the parameter space that allows for twin stars. Apart from these caveats our results should be applicable to a broad range of EoSs and quark matter parameterizations.
    \\\\
    Recent works have suggested that twin stars are unlikely to be compatible with astrophysical constraints but not entirely ruled out~\cite{Christian:2023hez,Christian:2025dhe,Grundler:2025mcz,Blomqvist:2025cxe}. Our analysis adds another argument against their realization if the phase transition occurs at high densities. However, phase transitions that do not produce twin stars are not strongly affected by our analysis, have been shown to fit astrophysical constraints \cite{Christian:2023hez,Christian:2025dhe} and appear likely to occur \cite{Annala:2023cwx,Tang:2025xib,Ayriyan:2025rub}.
	\begin{acknowledgments}
		JEC, PJ and SR were supported by the European Research Council (ERC) Advanced 
		Grant INSPIRATION under the European Union's Horizon 2020 research 
		and innovation programme (Grant agreement No. 101053985). SR was further supported by Deutsche Forschungsgemeinschaft (DFG, German Research Foundation) under Germany's Excellence Strategy - EXC 2121 "Quantum Universe" - 390833306, by the Swedish Research Council (VR) under grant number 2020-05044, by the research environment grant ``Gravitational Radiation and Electromagnetic Astrophysical Transients'' (GREAT) funded by the Swedish Research Council (VR) under Dnr 2016-06012 and  by the Knut-and-Alice-Wallenberg-Foundation under grant Dnr. KAW 2019.0112.
	\end{acknowledgments}
	
	\bibliographystyle{apsrev4-1}
	\bibliography{neue_bib_22,literature_eos}
\end{document}